\documentstyle [12pt]{article}
\begin{document}
\begin{titlepage}
\begin{flushright}
 IFUP-TH 43/94\\
 July 1994\\
\end{flushright}

\vspace{7mm}

\begin{center}

{\Large\bf Horizon quantization and thermodynamics\\

\vspace{2mm}

of the $2+1$ black hole}

\vspace{12mm}
{\large Michele Maggiore}

\vspace{3mm}

I.N.F.N. and Dipartimento di Fisica dell'Universit\`{a},\\
piazza Torricelli 2, I-56100 Pisa, Italy.\\
\end{center}

\vspace{4mm}

\begin{quote}
\hspace*{5mm} {\bf Abstract.} We test on the 2+1 dimensional black
hole the quantization method
that we have previously proposed in  four
dimensions. While in the latter case the horizon dynamics is
described, at the effective level, by
the action of a relativistic
membrane, in the 2+1 case it is described by a
string. We show that  this approach  reproduces the correct value
of the temperature and entropy of the 2+1 dimensional
black hole, and we write down
the Schroedinger equation satisfied by the horizon wave function.

\end{quote}
\end{titlepage}
\clearpage

In refs.~[1,2] we have proposed an approach for a
quantum description of black holes. The starting point is the idea
that, from the point of view of an observer external to the horizon and
static with respect to it, only
the degrees of freedom outside the horizon should enter as integration
variables in the path integral. This apparently simple requirement is
actually a very complicated and non-linear condition,
since the position of the horizon
is determined by the metric itself: when the metric fluctuates,
the horizon fluctuates and  the number
of variables $g_{\mu\nu}(x)$ which we would like to include
as integration variables
in the path integral changes. This non-linearity, however, becomes
tractable using a renormalization group (RG) approach in the spirit of
Wilson~\cite{Wil}: we first
introduce a fixed spherical surface located at $r=r_+
+\epsilon$, where $r_+$ is the radius of the classical horizon and
$\epsilon$ is larger than the
typical quantum fluctuations of the horizon,
i.e., a few Planck lengths. If $r>r_++\epsilon$ the
corresponding variables $g_{\mu\nu}$ are
inserted as integration variables
in the path integral. In the small shell between the fluctuating
horizon and the mathematical surface located at $r=r_++\epsilon$,
instead, we decompose the integration variables into
``collective coordinates''
$\zeta^{\mu}$
describing the location of the horizon,
plus ``fast variables", which are integrated over.
As discussed in~[2], the partition function
can therefore be written, neglecting the quantum fluctuations outside
the shell $r=r+\epsilon$, as
\begin{equation}\label{path2}
Z=\left( e^{- \mbox{\normalsize {\em I}}_{\rm grav} }
\int{\cal D}\zeta^{\mu}\,
e^{- \mbox{\normalsize {\em I}}_{\rm coll} }
\right) |_{g_{\mu\nu}=g_{\mu\nu}^{\rm cl}}\, ,
\end{equation}
where {\em I}$_{\rm grav}$ is the gravitational action including the
boundary term at $r=r_++\epsilon$ (which can be thought of as generated
by the RG procedure), and {\em I}$_{\rm coll}$ is the
effective action of the collective coordinates $\zeta^{\mu}$; it
is obtained integrating over the fast variables, and
its general form  is fixed by reparametrization
invariance. In the 3+1 case the simplest term which can arise is just
the action of a relativistic bosonic membrane.

Taking eq.~(\ref{path2}) as our basic starting point, we
are ready to compute a
number of quantities. To discuss thermal physics, we restrict the path
integral in eq.~(\ref{path2}) to configurations periodic in time, with
period $\beta_{\infty}$. (We reserve the notation $\beta$ for the
inverse of the local, blue-shifted temperature, see below.)
We have found  that, in 3+1 dimensions,
 there are periodic
solutions of the classical equations of motion of the Euclidean
membrane action. The inverse of the period of these configurations
gives the temperature at infinity
of the physical system. In Rindler space, denoting the acceleration by
$g$, we
have recovered in this way the
Unruh temperature $T=g/(2\pi)$, while for
Schwarzschild black holes (with mass $M$ above a critical value)
we have recovered the Hawking temperature,
$T=1/(8\pi M)$. The fact that this approach reproduces the correct
value of the temperature is by no means trivial. The derivation of
these results is  independent from the other ones
existing in the literature, as can also be seen from the fact that,
for black hole
masses below a critical value, the approach actually gives a
{\em different}
 prediction for the temperature! This latter result, togheter
with some appropriate qualifications, is discussed in~[2]. In this
paper we limit ourselves to the large mass region, which is
conceptually safer.

Having fixed $\beta_{\infty}$,
one can evaluate the partition function, and
hence the free energy and the entropy. The leading term is given by
the gravitational action in eq.~(\ref{path2}), and,
for Schwarzschild black holes in 3+1 dimensions,
 we have recovered
the $S={\rm area}/4$ relation.
Finally, we can discuss the quantization of the
action {\em I}$_{\rm coll}$ and, restricting
ourselves to radial modes
only, it is possible to derive a Schroedinger equation for the wave
function of the horizon, and to display its solution explicitly~[1].

This approach to quantum black holes is, in our opinion, a concrete
realization of general ideas proposed especially in
refs.~\cite{tH,Sus}. However, because of the coarse
graining over ``fast variables'', it also has
 the  limitations and the virtues of every
effective Lagrangian description: on the one hand, we cannot discuss
the fundamental problem of what happens at distances from the horizon
on the order of one Planck length, since we have performed, via the
renormalization group, a coarse graining over these scales. In order
to discuss short distances (and the related information loss paradox)
the fundamental theory is needed. On the other hand,
as far as larger scales are concerned, all the ignorance
about short distance physics is absorbed into a few
parameters, like the membrane tension, entering the action for the
collective coordinates. Furthermore, parameters  like the membrane
tension do not enter in the determination of the temperature
nor, to lowest order, of the entropy. They enter, instead, in the
computation of the (divergent) loop corrections to the entropy.

In order to examine the validity of the approach that we are proposing,
it is useful to discuss  its predictions in various physical
situations. As we have mentioned, in 3+1 dimensions the results are
encouraging:  the temperature and the entropy of black holes are
correctly reproduced.
The purpose of this Letter is to test
our approach in the ``theoretical
laboratory'' given by the
2+1 dimensional black hole solution discovered by
Ba\~nados, Teitelboim and Zanelli (BTZ)~\cite{BTZ}.
In spite of the many differences between three-dimensional and
four-dimensional gravity, the BTZ black hole has remarkable similarities
with its four-dimensional analog, and a number of investigations of
its geometrical and thermodynamical properties have appeared
recently, see e.g.~[6-9] and references therein.
One considers the theory defined by
the action
\begin{equation}\label{action}
I=\frac{1}{16\pi G}\int d^3x\, \sqrt{-g}\left[ R+2l^{-2}\right] +
B\, ,
\end{equation}
where $l$ is related to the cosmological constant $\Lambda$ by
$\Lambda =-l^{-2}<0$ and $B$ is the boundary term. Writing the metric
in the ADM form (and setting to zero the shift functions $N^i$),
$ds^2=N^2dt^2- g_{ij}dx^idx^j$,
the boundary term is given by
\begin{equation}
B=-\frac{1}{8\pi G}\int d^3x\,\partial_i\left[
g^{ij}\partial_jN\, \sqrt{^{(2)}g}\,\right]\, ,
\end{equation}
where $^{(2)}g={\rm det}\, g_{ij}$. Hereafter, following~\cite{BTZ}, we
will use units $G=1/8$.
The BTZ black hole, limiting ourselves to zero angular momentum, is
given by
\begin{equation}
ds^2=-\alpha dt^2 +\alpha^{-1} dr^2 +r^2d\theta^2\, ,
\hspace{7mm}
\alpha = \frac{r^2-r_+^2}{l^2}\, ,
\end{equation}
where $r_+=l\sqrt{M}$. In this case the
collective coordinates $\zeta^{\mu}$ depends on two variables
$\xi^i=(\tau
,\sigma )$ (we will later fix the gauge $\sigma =\theta$) and the
collective action is just the action of a
closed relativistic bosonic string
in $2+1$ dimensions
(plus higher order terms generated by the RG procedure,
which we neglect),
\begin{equation}
I_{\rm coll}=-{\cal T}\int d\sigma d\tau\,\sqrt{-h}\, .
\end{equation}
Here ${\cal T}$ is the string tension,
$h$ is the determinant of the induced metric
$h_{ij}=g_{\mu\nu}\partial_i\zeta^{\mu}\partial_j\zeta^{\nu}$,
 $x^{\mu}=\zeta^{\mu}(\xi )$ gives the
embedding of the string in 2+1 dimensional space-time and
$\partial_i = \partial /(\partial\xi^i)$;
$g_{\mu\nu}$ is the target space metric, eq.~(4).
We can now compute the temperature and entropy of the black hole
in our approach. The analysis closely parallels  the
$3+1$ case, so we only sketch the main points, referring the reader to
refs.~[1,2] for more detailed explanations.

Using the radial ansatz $r=r(\tau )$ and the gauge fixing
$x^{(0)}=\tau ,\theta =\sigma$, the equation of motion reads,
before rotating to Euclidean space,
\begin{equation}\label{eqm}
r\ddot{r}+(\alpha^2-\dot{r}^2)+\frac{r\alpha '}{2\alpha}
(\alpha^2-3\dot{r}^2)=0\, ,
\end{equation}
where  $\alpha '=d\alpha /dr$. Integrating, we get
$\dot{r}^2=\alpha^2-Cr^2\alpha^3$, with $C$ an integration
constant. This is very similar to the 3+1 case, and the qualitative
features of the motion are the same. The
circular string, depending on the
initial conditions, can either expand and then contract, or it
contracts immediately, and approaches the
nominal horizon asymptotically. The
equation of motion can be integrated exactly.
Defining $z=l\alpha^{1/2}=(r^2-r_+^2)^{1/2}$
and $g=r_+/l^2=\sqrt{M}/l$, the equation of motion for $z$ reads
\begin{equation}\label{11}
z\ddot{z}-2\dot{z}^2+g^2z^2 +(\frac{2g^2}{r_+^2})z^4
-2\frac{z^2\dot{z}^2}{r_+^2+z^2}=0\, .
\end{equation}
With the initial conditions $z(0)=ar_+,\dot{z}(0)=0$, where $a$ is a
dimensionless parameter, the exact solution is
\begin{equation}
g\tau = \frac{1}{a^2\sqrt{1+2 a^2}}\, \Pi (\psi ,\frac{1+a^2}{a^2},
\sqrt{\frac{1+a^2}{1+2 a^2}})\, .
\end{equation}
Here $\Pi (\psi ,\alpha^2,k)$ is the elliptic integral of the third
kind, and $\sin^2\psi = (a^2r_+^2-z^2)/(a^2r_+^2+r_+^2)$.

The typical fluctuations of the horizon are on the order of a few
Planck length, so that, as far as the dependence on the
black hole mass $M$ is
concerned,
 $z(0)\sim 1$ and therefore $a\sim M^{-1/2}$ (while in $3+1$ we got
$a\sim M^{-1}$). Thus, for  large black hole masses, the amplitude is
small and we can neglect the non-linear terms in eq.~(\ref{11}), which
becomes $z\ddot{z}-2\dot{z}^2+g^2z^2 \simeq 0$. This is the same
equation that we found in~[1,2]; after rotation to Euclidean space,
$\tau\rightarrow i\tau$, it has a periodic solution with period
$\beta_{\infty} = 2\pi/g$ and therefore the temperature is
\begin{equation}
T=\frac{g}{2\pi}=\frac{\sqrt{M}}{2\pi l}\, ,
\end{equation}
which agrees with the result found in ref.~\cite{BTZ}.

Having fixed
$\beta_{\infty}$, we can  compute
the entropy of the  black
hole, evaluating the partition function $Z$, and therefore the free
energy.  The leading term is given by the gravitational action, while
the string action gives the loop corrections. The volume term in
$I_{\rm grav}$ just contributes to the zero point of the free energy,
so the only non trivial contribution comes from the
boundary term evaluated on the surface $r=r_++\epsilon$.
To compute the entropy $S$ we follow the method suggested by
York~\cite{Yo86}: we define the local inverse temperature
$\beta =\beta_{\infty}\alpha^{1/2}$ and we compute
the boundary term on a
spherical surface with a generic radius $r$. Considering $M$ as a
function of $\beta$ and $r$ defined implicitly by
$\beta =\beta_{\infty}\alpha^{1/2}$, we find $B=B(\beta ,r)$, and the
entropy is given by
\begin{equation}
S=\beta\left(\frac{\partial B}{\partial\beta}\right)_r-B\, .
\end{equation}
A simple computation gives
\begin{equation}
B=-2\beta_{\infty}\frac{r^2}{l^2}=-4\pi r
\left[ 1+(\frac{\beta}{2\pi l})^2\right]^{1/2}\, ,
\end{equation}
and therefore
\begin{equation}
S=4\pi r_+ = 2\times {\rm perimeter}\, ,
\end{equation}
in agreement with ref.~\cite{BTZ}.\footnote{In ref.~\cite{ER} units
$G=1$ are used,
and therefore the result reads $S=({\rm perimeter})/4.$}
Note that the result for $S$ is independent of $r$,
the radius
of the surface on which the boundary term is computed.
Thus,  the dependence
on the exact position of the surface
used in the RG procedure disappears.

We conclude that our approach reproduces correctly the thermodynamical
properties of the BTZ black hole.

Finally, we discuss the quantization of the black hole horizon along
the lines of ref.~[1], i.e. limiting ourselves to
a  minisuperspace approximation in which only spherical
membranes, $r=r(\tau )$, are considered.
The effective lagrangian for the radial mode is obtained inserting
$r=r(\tau )$, togheter with the gauge fixing $x^{(0)}=\tau,\theta
=\sigma$, into the membrane action. The result is
\begin{equation}
L=-2\pi{\cal T}\, r(\alpha-\frac{\dot{r}^2}{\alpha})^{1/2}\, .
\end{equation}
The equation of motion of this lagrangian is just eq.~(\ref{eqm}).
We introduce the tortoise coordinate $r_*$, defined by
$dr_*=dr/\alpha$, or
\begin{equation}
r_*=\frac{l^2}{2r_+}\log\frac{r-r_+}{r+r_+}\, .
\end{equation}
As $r$ ranges between $r_+$ and $\infty$, $r_*$ ranges between
$-\infty$ and zero.
The relation between $r_*$ and $r$ can be inverted analytically, and
gives
\begin{equation}
r=-r_+\coth (r_+r_*/l^2)\, .
\end{equation}
Defining the momentum conjugate to
$r_*$, $p_*=\delta L/(\delta \dot{r}_*)$, we get the Hamiltonian
\begin{equation}
H=p_*\dot{r}_*-L=\sqrt{p_*^2+(2\pi{\cal T})^2r^2\alpha}\, ,
\end{equation}
and then the Schroedinger equation for the wave function of the
horizon,
\begin{equation}
\left( -\frac{d^2}{dx^2}+\mu^2\, \frac{\cosh^2 x}{\sinh^4 x}
\right)\psi(x)=\varepsilon\psi(x)\, ,
\end{equation}
where $x=r_+r_*/l^2$ ranges between $-\infty$ and zero,
 $\mu = 2\pi{\cal T}r_+l$, and $\varepsilon$ is related to the
excitation energy of the membrane $E$ by
$\varepsilon =l^4E^2/r_+^2$. The qualitative features of this
Schroedinger equation are absolutely identical to the $3+1$ case, and
we refer the reader to ref.~[1] for a discussion of its physical
meaning.


\begin{thebibliography}{999}
\bibitem{MM1} M. Maggiore, {\em ``Black Holes as Quantum Membranes''},
preprint IFUP-TH 6/94, January 1994, gr-qc/9401027.
\bibitem{MM2} M. Maggiore, {\em ``Black holes as quantum membranes:
path integral approach''},
preprint IFUP-TH 22/94, April 1994, hep-th/9404172,
{ Phys. Lett.}  B (to be published).
\bibitem{Wil} K.G. Wilson, Rev. Mod. Phys. 55 (1983) 583;

K.G. Wilson and J.~Kogut, Phys. Rep. 12 (1974) 75.

\bibitem{tH} G. 't Hooft,  { Nucl. Phys.}  B256 (1985) 727,
{ Nucl. Phys.}  B335 (1990) 138;
Physica Scripta T36 (1991) 247.
\bibitem{Sus} L. Susskind, L. Thorlacius and J.~Uglum,
{ Phys. Rev.}  D48 (1993) 3743;

L. Susskind, {Phys. Rev. Lett.}  71 (1993) 2367;
{ Phys. Rev.}  D49 (1994) 6606;

L. Susskind and L. Thorlacius, { Phys. Rev.}  D49 (1994) 966.
\bibitem{BTZ} M. Ba\~nados, C.~Teitelboim and J.~Zanelli,
{Phys. Rev. Lett.}  69 (1992) 1849.
\bibitem{BTHZ} M. Ba\~nados, M.~Henneaux,
C.~Teitelboim and J.~Zanelli, { Phys. Rev.}  D48 (1993) 1506.
\bibitem{ER} F.~Englert
and B.~Reznik, {\em ``Entropy Generation by Tunneling
in 2+1-Gravity''}, preprint TAUP-2102-93, gr-qc/9401010.
\bibitem{Rez} B. Reznik,
{\em ``Thermodynamics and Evaporation of the 2+1-D
Black Hole''}, preprint TAUP 2141/94, gr-qc/9403027.
\bibitem{Yo86} J.W. York, Jr., { Phys. Rev.}  D33 (1986) 2092.
\end{thebibliography}
\end{document}